\documentclass[prd,nofootinbib,showpacs,preprint]{revtex4}
\usepackage{amsmath}
\usepackage{graphicx}
\graphicspath{{Figs/}}
\usepackage{dcolumn}
\usepackage{bm}
\usepackage{amssymb}
\usepackage[usenames,dvipsnames]{color}
\usepackage{slashed}
\usepackage[dvipdfm,colorlinks,citecolor=blue]{hyperref}
\begin{document}
\title{Deviations from Tri-Bimaximal Neutrino Mixing \\ Using Type II Seesaw}
\author{Debasish Borah}
\email{dborah@tezu.ernet.in}
\affiliation{Department of Physics, Tezpur University, Tezpur-784028, India}


\begin{abstract}
We study the possibility of generating deviations from tri-bimaximal (TBM) neutrino mixing to explain the non-zero reactor mixing angle within the framework of both type I and type II seesaw mechanisms. The type I seesaw term gives rise to the $\mu-\tau$ symmetric TBM pattern of neutrino mass matrix as predicted by generic flavor symmetry models like $A_4$ whereas the type II seesaw term gives rise to the required deviations from TBM pattern to explain the non-zero $\theta_{13}$. Considering extremal values of Majorana CP phases such that the neutrino mass eigenvalues have the structure $(m_1, -m_2, m_3)$ and $(m_1, m_2, m_3)$, we numerically fit the type I seesaw term by taking oscillation as well as cosmology data and then compute the predictions for neutrino parameters after the type II seesaw term is introduced. We consider a minimal structure of the type II seesaw term and check whether the predictions for neutrino parameters lie in the $3\sigma$ range. We also outline two possible flavor symmetry models to justify the minimal structure of the type II seesaw term considered in the analysis. 
\end{abstract}
\pacs{12.60.-i,12.60.Cn,14.60.Pq}
\maketitle
\section{Introduction}
The existence of the non-zero yet tiny neutrino masses \cite{PDG} have been confirmed by several neutrino oscillation experiments in the last few years. The smallness (relative to the charged fermion masses) of three Standard Model
neutrino masses can be naturally explained 
via seesaw mechanism which can be of three types : type I \cite{ti}, type II \cite{tii} and type III \cite{tiii}. All these mechanisms involve the inclusion of additional fermionic or scalar fields to generate tiny neutrino masses at tree level. Although these seesaw models can naturally explain the smallness of neutrino mass compared to the electroweak scale, we are still far away from understanding the origin of neutrino mass hierarchies as suggested by experiments. Recent neutrino oscillation experiments T2K \cite{T2K}, Double ChooZ \cite{chooz}, Daya-Bay \cite{daya} and RENO \cite{reno} have not only made the earlier predictions for neutrino parameters more precise, but also predicted non-zero value of the reactor mixing angle $\theta_{13}$. The latest global fit value for $3\sigma$ range of neutrino oscillation parameters \cite{schwetz12} are as follows:
$$ \Delta m_{21}^2=(7.00-8.09) \times 10^{-5} \; \text{eV}^2$$
$$ \Delta m_{31}^2 \;(\text{NH}) =(2.27-2.69)\times 10^{-3} \; \text{eV}^2 $$
$$ \Delta m_{23}^2 \;(\text{IH}) =(2.24-2.65)\times 10^{-3} \; \text{eV}^2 $$
$$ \text{sin}^2\theta_{12}=0.27-0.34 $$
$$ \text{sin}^2\theta_{23}=0.34-0.67 $$ 
\begin{equation}
\text{sin}^2\theta_{13}=0.016-0.030
\end{equation}
where NH and IH refers to normal and inverted hierarchy respectively. The best fit value of $\delta_{CP}$ turns out to be $300$ degrees \cite{schwetz12}.

Prior to the discovery of non-zero $\theta_{13}$, the neutrino oscillation data were compatible with the so called TBM form of the neutrino mixing matrix discussed extensively in the literature\cite{Harrison} which is given by
\begin{equation}
U_{TBM}==\left(\begin{array}{ccc}\sqrt{\frac{2}{3}}&\frac{1}{\sqrt{3}}&0\\
 -\frac{1}{\sqrt{6}}&\frac{1}{\sqrt{3}}&\frac{1}{\sqrt{2}}\\
\frac{1}{\sqrt{6}}&-\frac{1}{\sqrt{3}}& \frac{1}{\sqrt{2}}\end{array}\right),
\end{equation}
which predicts $\text{sin}^2\theta_{12}=\frac{1}{3}$, $\text{sin}^2\theta_{23}=\frac{1}{2}$ and $\text{sin}^2\theta_{13}=0$. However, since the latest data have ruled out $\text{sin}^2\theta_{13}=0$, there arises the need to go beyond the TBM framework. Since the experimental value of $\theta_{13}$ is still much smaller than the other two mixing angles, TBM can still be a valid approximation and the non-zero $\theta_{13}$ can be accounted for by incorporating non-leading contributions to TBM coming from charged lepton mass diagonalization, for example. There have already been a great deal of activities in this context \cite{nzt13, nzt13GA} and the latest data can be successfully predicted within the framework of several interesting models. These frameworks which predict non-zero $\theta_{13}$ may also shed light on the Dirac CP violating phase which is still unknown (and could have remained unknown if $\theta_{13}$ were exactly zero).

Apart from predicting the correct neutrino oscillation data as well as the Dirac CP phase, the nature of neutrino mass hierarchy is also an important yet unresolved issue. Understanding the correct nature of hierarchy can also have non-trivial relevance in leptogenesis as well as cosmology. Recently such a comparative study was done to understand the impact of mass hierarchies, Dirac and Majorana CP phases on the predictions for baryon asymmetry in \cite{leptodborah}. Another cosmologically relevant parameter in neutrino physics is the sum of absolute neutrino masses. The constraint on the sum of absolute neutrino masses from largest photometric redshift survey \cite{SFO} have already ruled out the scenario of quasi-degenerate (QDN) neutrino masses with $m_i \geq 0.1 \; \text{eV}$. From supernova neutrinos point of view, it was shown \cite{HM} that one can discriminate the inverted hierarchy from the normal one if $\text{sin}^2\theta_{13}\geq \text{a few} \times 10^{-4}$. Recently the Planck collaboration has reported a more stringent bound on the sum of absolute neutrino masses $\sum_i m_{i} < 0.23$ eV \cite{Planck13}. Therefore, the study of normal and inverted hierarchy using different types of seesaw mechanism is very important both from neutrino physics and cosmology point of view.

In view of above, the present work is planned to study the possibility of generating correct oscillation and cosmology data by considering type I seesaw term giving rise to TBM type neutrino mixing and type II seesaw term as a perturbation to the leading order TBM mixing. Such a work was done recently in \cite{mkd-db-rm} 
where either type I or type II term was considered as leading order and the impact of the other term as a small perturbation on neutrino parameters was studied. In another work \cite{dbgrav}, the impact of Planck suppressed operators on neutrino mixing parameters was studied. In these works, the leading order contribution to neutrino masses was fitted with the global best-fit neutrino data (including non-zero $\theta_{13}$) and then the effects of non-leading contributions were studied. In the present work, however, we assume the leading contribution to neutrino mass (the type I seesaw term) as TBM type which is numerically fitted with the oscillation data on mass squared differences and cosmological upper bound on the sum of absolute neutrino masses. The motivation behind this assumption is the dynamical origin of TBM mixing pattern in terms of a broken flavor symmetry based on discrete groups like $A_4$ \cite{A4TBM}. The type II seesaw term is then introduced as a perturbation and the predictions for the neutrino parameters are calculated. Similar attempts to study the deviations from TBM mixing by using the interplay of two different seesaw mechanisms were done in \cite{devtbmt2}. Our work is different in the sense that we have considered type I seesaw term as the leading order contribution giving rise to TBM mixing and type II seesaw term as the perturbation which generates the required deviation from TBM mixing in order to explain the non-zero reactor mixing angle. We vary the strength of this perturbation and check whether the same strength of the perturbation can generate non-zero $\theta_{13}$ in agreement with experiments and also keep the other neutrino parameters within the $3\sigma$ range of global fit data. We also check whether the sum of absolute neutrino masses obey the cosmological upper bound as we vary the strength of the perturbation. We consider both normal and inverted hierarchical neutrino mass patterns as well as two different Majorana CP phase patterns ($(m_1,-m_2,m_3)$ and $(m_1,m_2,m_3)$) denoted by $(+-+)$ and $(+++)$ respectively. Out of these four cases we study, only inverted hierarchical case with $(+++)$ type Majorana phases has all the parameters within or very close to the $3\sigma$ range whereas for other cases, at least one of the parameters fall far outside the $3\sigma$ range. For the numerical analysis we stick to a minimal structure of the type II seesaw term which we justify by a brief outline of two flavor symmetry models. Similar works were carried out

This paper is organized as follows: in section \ref{method} we discuss the methodology of type I and type II seesaw mechanisms. In section \ref{devTBM}, we discuss the parametrization of TBM type $\mu-\tau$ symmetric neutrino mass matrix as well as the deviations from TBM mixing in order to generate non-zero reactor mixing angle. In section \ref{flav}, we outline two different flavor symmetry models that can explain the minimal structure of the type II seesaw term we adopt in our analysis. In section \ref{numeric} we discuss our numerical analysis and results and then finally conclude in section \ref{conclude}.
\section{Seesaw Mechanism: Type I and Type II}
\label{method}
Seesaw mechanism is a natural way to explain the origin of tiny $(~eV)$ neutrino masses. Type I seesaw \cite{ti} framework is the simplest of them where the standard model is extended by inclusion of three right handed neutrinos $(\nu^i_R, i = 1,2,3)$ as gauge singlets. Being singlet under the gauge group, bare mass terms of the right handed neutrinos $M_{RR}$ are allowed. The resulting type I seesaw formula  is given by the expression,
\begin{equation}
m_{LL}^I=-m_{LR}M_{RR}^{-1}m_{LR}^{T}.
\end{equation}
where $m_{LR}$ is the Dirac mass term of the neutrinos. Thus, if $M_{RR}$ is as high as $10^{14}$ GeV, tiny neutrino mass can be explained naturally without any fine-tuning of Dirac Yukawa couplings. 

In type II seesaw \cite{tii} mechanism, the standard model is extended by inclusion of a Higgs field which is triplet under $SU(2))_L$ as follows:
\begin{equation}
\Delta_L =
\left(\begin{array}{cc}
\ \delta^+_L/\surd 2 & \delta^{++}_L \\
\ \delta^0_L & -\delta^+_L/\surd 2
\end{array}\right) \nonumber
\end{equation} 
This allows the term in the Yukawa Lagrangian
$ f_{ij}\ \left(\ell_{iL}^T \ C \ i \sigma_2 \Delta_L \ell_{jL}\right)$ which can account for tiny neutrino mass if the neutral component of the Higgs triplet $\delta^0_L$ acquires a tiny vacuum expectation value (vev). From the minimization of the scalar potential, it turns out that the vev of $\delta^0_L$ is given by 
\begin{equation}
 \langle \delta^0_L \rangle = v_L = \frac{\langle \phi^0 \rangle^2}{M_{\Delta}}
\label{vev} 
\end{equation}
where $\phi^0$ is the neutral component of the electroweak Higgs doublet with vev $~10^2$ GeV. Thus, $M_{\Delta}$ has to be as high as $10^{14}$ GeV to give rise to tiny neutrino masses without any fine-tuning of the couplings $f_{ij}$.
\begin{table}
\centering
\caption{Parametrization of the neutrino mass matrix for TBM mixing}
\vspace{0.5cm}
{\small
\begin{tabular}{|c|c|c|c|c|}
 \hline
   Parameters & IH(+-+) &  IH(+++) &  NH(+-+)&  NH(+++)\\ \hline
x&0.0685152&0.0685152&0.0206981&0.0628882\\  \hline
y&0.00550731&0.00550731&-0.0419916&0.000198452\\  \hline
z&-0.071504&0.00849603&0.00865514&0.00865514\\  \hline
$m_3\;(\text{eV})$&0.0630079&0.0630079&0.08&0.08\\  \hline
$m_2\; (\text{eV})$&-0.08&0.08&-0.0632851&0.0632851\\  \hline
$m_1\;(\text{eV})$&0.0795298&0.0795298&0.0626897&0.0626897\\  \hline
$\sum_i m_i\;(\text{eV})$&0.2225&0.2225&0.2059&0.2059\\  \hline
\end{tabular}
}
\label{table:results1}
\end{table}
\section{Deviations from TBM mixing}
\label{devTBM}
Type I seesaw giving rise to $\mu-\tau$ symmetric TBM mixing pattern for neutrinos have been discussed extensively in the literature. The neutrino mass matrix in these scenarios can be parametrized as
\begin{equation}
m_{LL}=\left(\begin{array}{ccc}
x& y&y\\
y& x+z & y-z \\
y & y-z & x+z
\end{array}\right)
\label{matrix1}
\end{equation}
which is clearly $\mu-\tau$ symmetric with eigenvalues $m_1 = x-y, \; m_2 = x+2y, \; m_3 = x-y+2z$. It predicts the mixing angles as $\theta_{12} \simeq 35.3^o, \; \theta_{23} = 45^o$ and $\theta_{13} = 0$. Although the prediction for first two mixing angles are still allowed from oscillation data, $\theta_{13}=0$ has been ruled out experimentally at more than $9\sigma$ confidence level. This has led to a significant number of interesting works trying to explain the origin of non-zero $\theta_{13}$. Here we study the possibility of explaining the deviations from TBM mixing and hence from $\theta_{13}=0$ by allowing the type II seesaw term as a perturbation.

Before choosing the minimal structure of the type II seesaw term, we note that the parametrization of the TBM plus corrected neutrino mass matrix can be done as \cite{nzt13GA}.
\begin{equation}
m_{LL}=\left(\begin{array}{ccc}
x& y-w&y+w\\
y-w& x+z+w & y-z \\
y+w & y-z& x+z-w
\end{array}\right)
\label{matrix2}
\end{equation}
where $w$ denotes the deviation of $m_{LL}$ from that within TBM frameworks and setting it to zero, the above matrix boils down to the familiar $\mu-\tau$ symmetric matrix (\ref{matrix1}). Thus, the minimal structure of the perturbation term to the leading order $\mu-\tau$ symmetric TBM neutrino mass matrix can be taken as
\begin{equation}
m^{II}_{LL}=\left(\begin{array}{ccc}
0& -w& w\\
-w& w & 0 \\
w & 0& -w
\end{array}\right)
\label{matrix3}
\end{equation}
Such a stucture of the type II seesaw term can be explained by continuous as well as discrete flavor symmetries as we briefly outline in the next section.

\section{Flavor Symmetry Explanation of $m^{II}_{LL}$}
\label{flav}
The minimal structure of the type II seesaw term (\ref{matrix3}) we adopt here can be explained by incorporating the presence of flavor symmetries. Here we briefly outline two possibilities: one within the framework of abelian gauged flavor symmetry and the other within the framework of $A_4$ flavor symmetry. For simplicity of our analysis, we assume that these additional symmetries still allow the type I seesaw term to give rise to the $\mu-\tau$ symmetric TBM neutrino mass matrix (\ref{matrix1}).

\subsection{Abelian Flavor Symmetry}
Abelian gauge extension of Standard Model is one of the best motivating examples of beyond Standard Model physics. 
For a review see \cite{Langacker}. Such a model is also motivated within the framework of GUT models, for example $E_6$. 
The supersymmetric version of such models have an additional advantage in the sense that they provide a solution to the MSSM (Minimal Supersymmetric Standard Model) $\mu$ problem. Such abelian gauge extension of SM was studied recently in \cite{Borah} in the context of neutrino mass and cosmology.

Here we consider an extension of the Standard Model gauge group with one abelian $U(1)_X$ gauge symmetry. Thus, the model we propose here is an $SU(3)_c \times SU(2)_L \times U(1)_Y \times U(1)_X$ gauge theory with three chiral generations of SM and three additional right handed neutrinos. We will consider family 
non universal $U(1)_X$ couplings.

The fermion content of our model is 
\begin{equation}
Q_i=
\left(\begin{array}{c}
\ u \\
\ d
\end{array}\right)
\sim (3,2,\frac{1}{6},n_{qi}),\hspace*{0.8cm}
L_i=
\left(\begin{array}{c}
\ \nu \\
\ e
\end{array}\right)
\sim (1,2,-\frac{1}{2},n_{li}), \nonumber 
\end{equation}
\begin{equation}
u^c_i \sim (3^*,1,\frac{2}{3},n_{ui}), \quad d^c_i \sim (3^*,1,-\frac{1}{3},n_{di}), \quad e^c_i \sim (1,1,-1,n_{ei}), \quad \nu^c_i \sim (1,1,0,n_{ri}) \nonumber 
\end{equation}
where $ i=1,2,3 $ goes over the three generations of Standard Model and the numbers in the bracket correspond to the quantum number under the gauge group $SU(3)_c \times SU(2)_L \times U(1)_Y \times U(1)_X$. The $U(1)_X$ gauge quantum numbers should be such that they do not give rise to anomalies. We consider the following solution of the anomaly matching conditions
$$ n_{qi}=n_{ui}=n_{di}=0, \;\;\; n_{li}=n_{ei} = n_{ri} = n_i $$
$$ \sum n_{li} = \sum n_{ei} = \sum n_{ri} = 0, \;\;\;  \sum n^3_{li} = \sum n^3_{ei} = \sum n^3_{ri} = 0$$
In particular, we choose $n_1 = 0, n_2 = n, n_3 = -n$. Thus to have the desired structure of the type II seesaw term shown in (\ref{matrix3}), we need four Higgs triplet fields $\Delta_{1,2,3,4}$ with gauge quantum numbers $(1,3,1,n), (1,3,1,-n), (1,3,1,2n)$ and $(1,3,1,-2n)$ respectively. It can be noticed that due to non-universal gauge quantum numbers of the fermions and the existence of four triplet fields, the type II seesaw term $f_{ij} L_i L_j \Delta$ is gauge invariant only for $f_{12}L_1 L_2 \Delta_2, f_{13}L_1L_3 \Delta_1, f_{22} L_2L_2 \Delta_4, f_{33} L_3L_3\Delta_3 $ combinations. Hence, the type II contribution to neutrino mass matrix will have zero entries in the $(1,1), (2,3)$ and $(3,2)$ components as in the matrix (\ref{matrix3}). Although the abelian flavor symmetry can provide an explanation for the zero entries in the matrix, it does not dictate any relations between the non-zero parameters which, in general, can be different from each other. For the minimal choice we adopt in our discussion, we further assume that the non-zero entries are different only upto a sign.
\begin{figure}[ht]
 \centering
\includegraphics[width=1.0\textwidth]{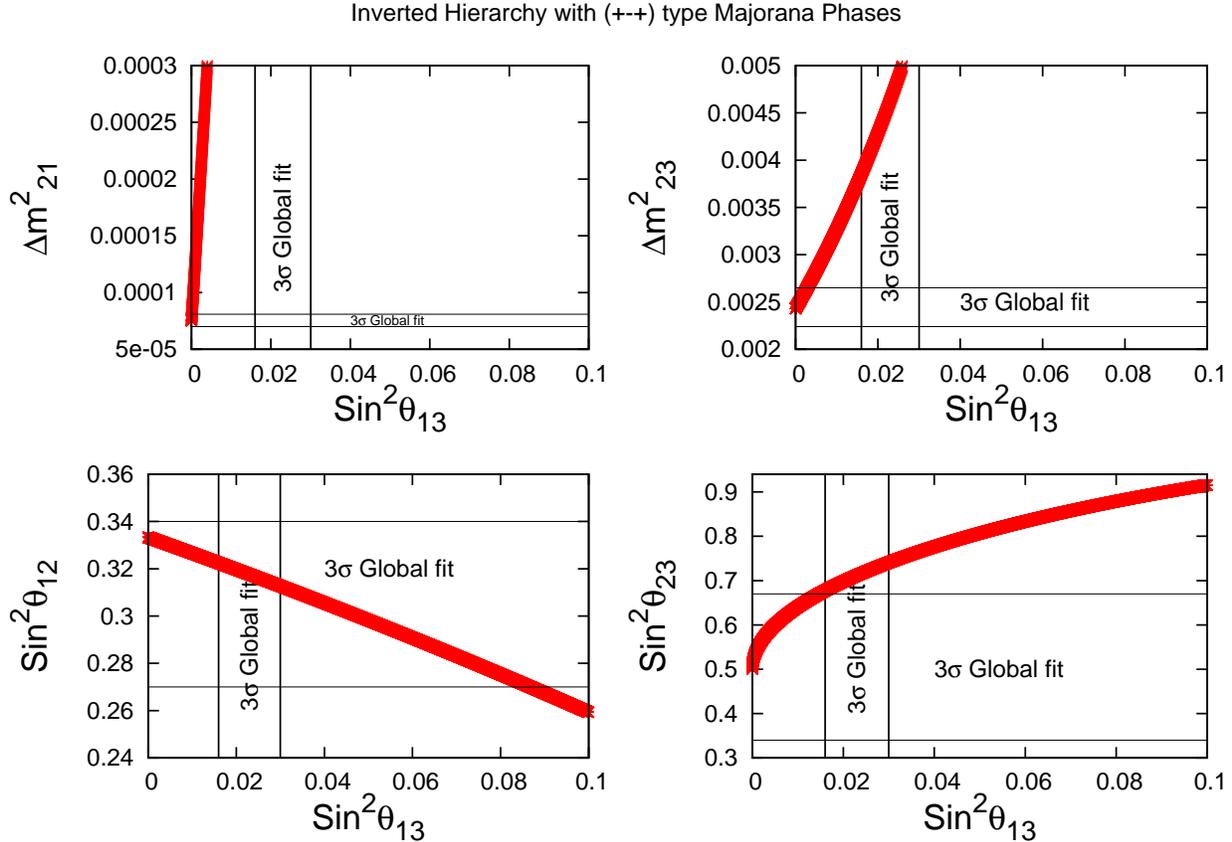}
\caption{Variation of neutrino parameters as a function of $\sin^2{\theta_{13}}$ for $IH(+-+)$}
\label{fig1}
\end{figure}
\begin{figure}[ht]
 \centering
\includegraphics[width=1.0\textwidth]{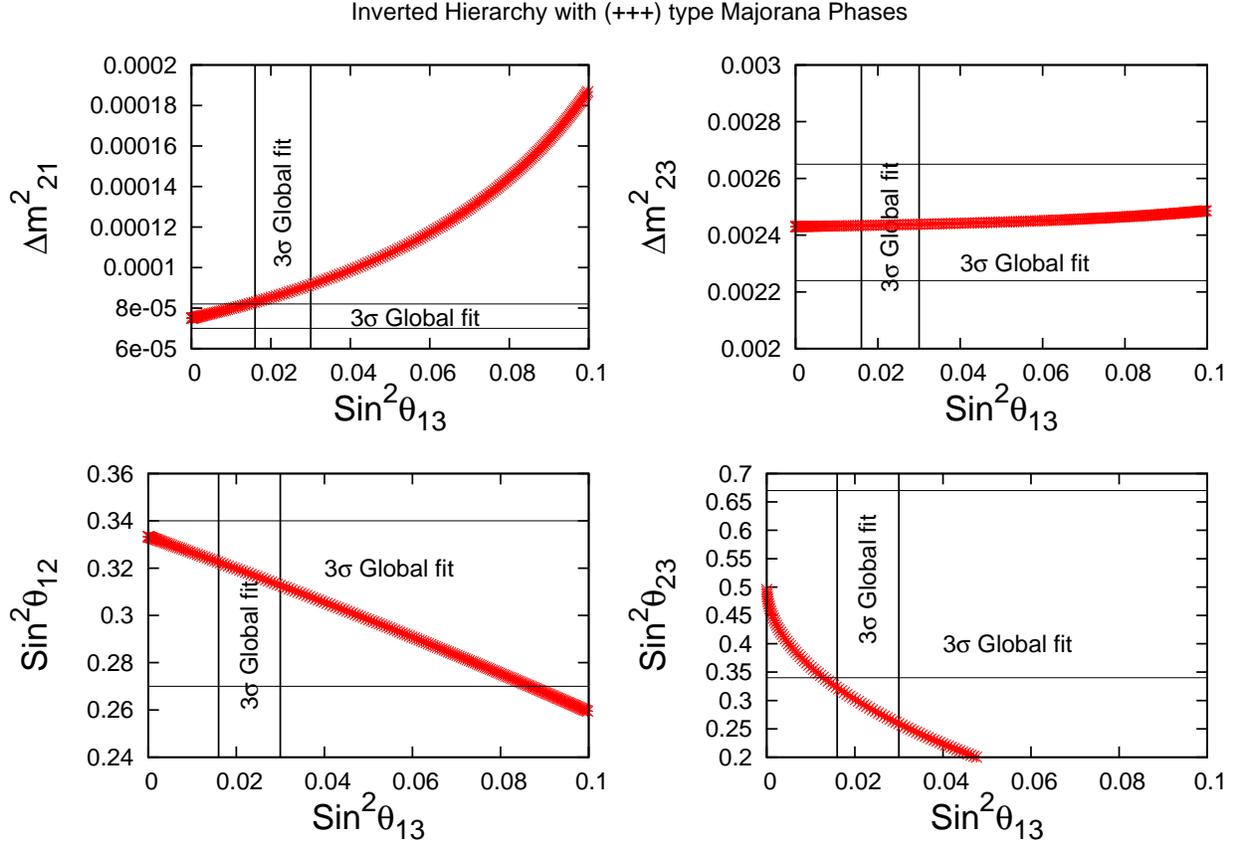}
\caption{Variation of neutrino parameters as a function of $\sin^2{\theta_{13}}$ for $IH(+++)$}
\label{fig2}
\end{figure}
\subsection{$A_4$ flavor symmetry} 
The discrete group $A_4$ consists of the even permutations of four objects. It is also the symmetry group of a tetrahedron. This is one of the most popular group in the discrete flavor symmetry literature which can naturally explain the TBM pattern of neutrino mixing \cite{A4TBM}, a good approximation to neutrino oscillation data at leading order. Non-leading corrections to TBM within the context of $A_4$ symmetric model has also been addressed in some of the works mentioned in references \cite{nzt13, nzt13GA}. Here we do not discuss the realization of TBM within $A_4$ symmetry for type I seesaw model. This has been addressed several times in the literature. We only outline the dynamical realization of a specific minimal structure of the type II seesaw term which we use in our numerical analysis. 

The group $A_4$ has four irreducible representations namely, $\bf{1}, \bf{1'}, \bf{1"}$ and $\bf{3}$. In generic $A_4$ models, the $SU(2)_L$ lepton doublets $l = (l_e, l_{\mu}, l_{\tau})$ are assumed to transform as triplet $\bf{3}$ under $A_4$ whereas the $SU(2)_L$ singlet charged leptons $e^c, \mu^c, \tau^c$ transform as $\bf{1}, \bf{1'}, \bf{1"}$ respectively. In type I seesaw scenarios, the $SU(2)_L$ singlet right handed neutrinos $\nu^c$ transform as a triplet under $A_4$. Since we are trying to explain the structure of type II term only, we confine our discussion to the lepton doublets only. We introduce an $A_4$ triplet scalar field $\zeta$ which is chargeless under the standard model gauge symmetry. The $SU(2)_L$ triplet Higgs field $\Delta_L$ is assumed to be a singlet under $A_4$. We also incorporate an additional $Z_2$ symmetry under which $\zeta \rightarrow -\zeta,\; \Delta_L \rightarrow -\Delta_L$ whereas all other fields are even under it. This also prevents the scalar field $\zeta$ to couple to the type I seesaw term. Thus the type II seesaw term can be written as 
$$ \mathcal{L}^{II} = f l l \zeta \Delta_L/\Lambda$$
where $\Lambda$ is the cutoff scale and $f$ is a dimensionless coupling constant. Decomposition of $ll\zeta$ term into an $A_4$ singlet gives 
$$ ll\zeta \rightarrow (2l_el_e-l_{\mu}l_{\tau}-l_{\tau}l_{\mu})\zeta_1+(2l_{\tau}l_{\tau}-l_el_{\mu}-l_{\mu}l_e)\zeta_2+(2l_{\mu}l_{\mu}-l_el_{\tau}-l_{\tau}l_e)\zeta_3 $$
Assuming the vacuum alignment of the $A_4$ triplet field as $\langle \zeta \rangle = \alpha \Lambda (0,-1,1)$, we obtain the type II seesaw contribution to neutrino mass as
\begin{equation}
m^{II}_{LL}=\left(\begin{array}{ccc}
0& w& -w\\
w& 2w & 0 \\
-w & 0& -2w
\end{array}\right)
\label{matrix4}
\end{equation}
where $w = f \langle \delta^0_L \rangle = fv_L$. We adopt this minimal structure of the type II seesaw term for our numerical analysis in the next section.

\section{Numerical Analysis and Results}
\label{numeric}
For our numerical analysis, we adopt the minimal structure (\ref{matrix4}) of the type II seesaw term dictated by $A_4$ flavor symmetry. We first numerically fit the leading order $\mu-\tau$ symmetric neutrino mass matrix (\ref{matrix1}) by taking the central values of the global fit neutrino oscillation data. We also incorporate the cosmological upper bound on the sum of absolute neutrino masses reported by the Planck collaboration recently. We consider both normal and inverted hierarchical neutrino masses as well as both the possibilities of extremal Majorana neutrino phases $(+++), (+-+)$. The parametrization for all these possible cases are shown in table \ref{table:results1}.

\begin{figure}[ht]
 \centering
\includegraphics[width=1.0\textwidth]{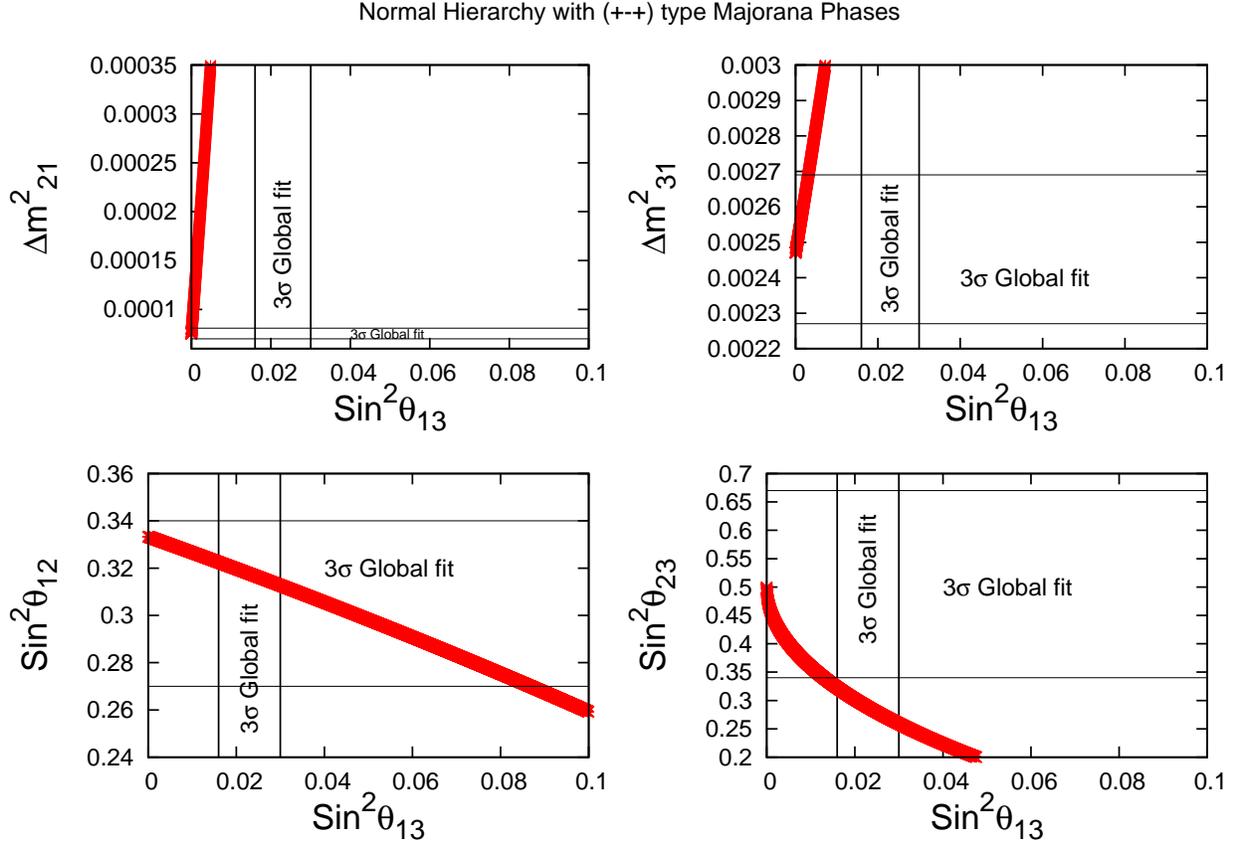}
\caption{Variation of neutrino parameters as a function of $\sin^2{\theta_{13}}$ for $NH(+-+)$}
\label{fig3}
\end{figure}
\begin{figure}[ht]
 \centering
\includegraphics[width=1.0\textwidth]{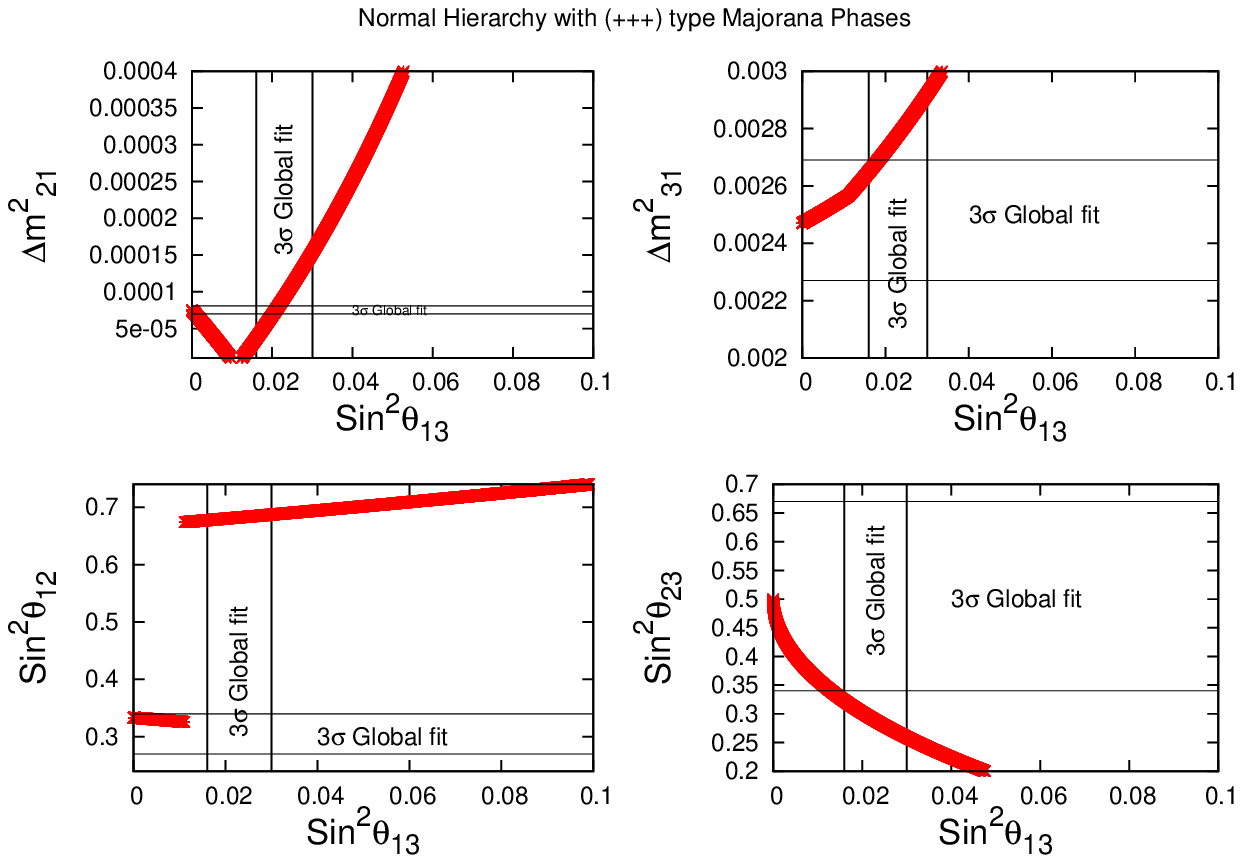}
\caption{Variation of neutrino parameters as a function of $\sin^2{\theta_{13}}$ for $NH(+++)$}
\label{fig4}
\end{figure}
\begin{figure}[ht]
 \centering
\includegraphics[width=1.0\textwidth]{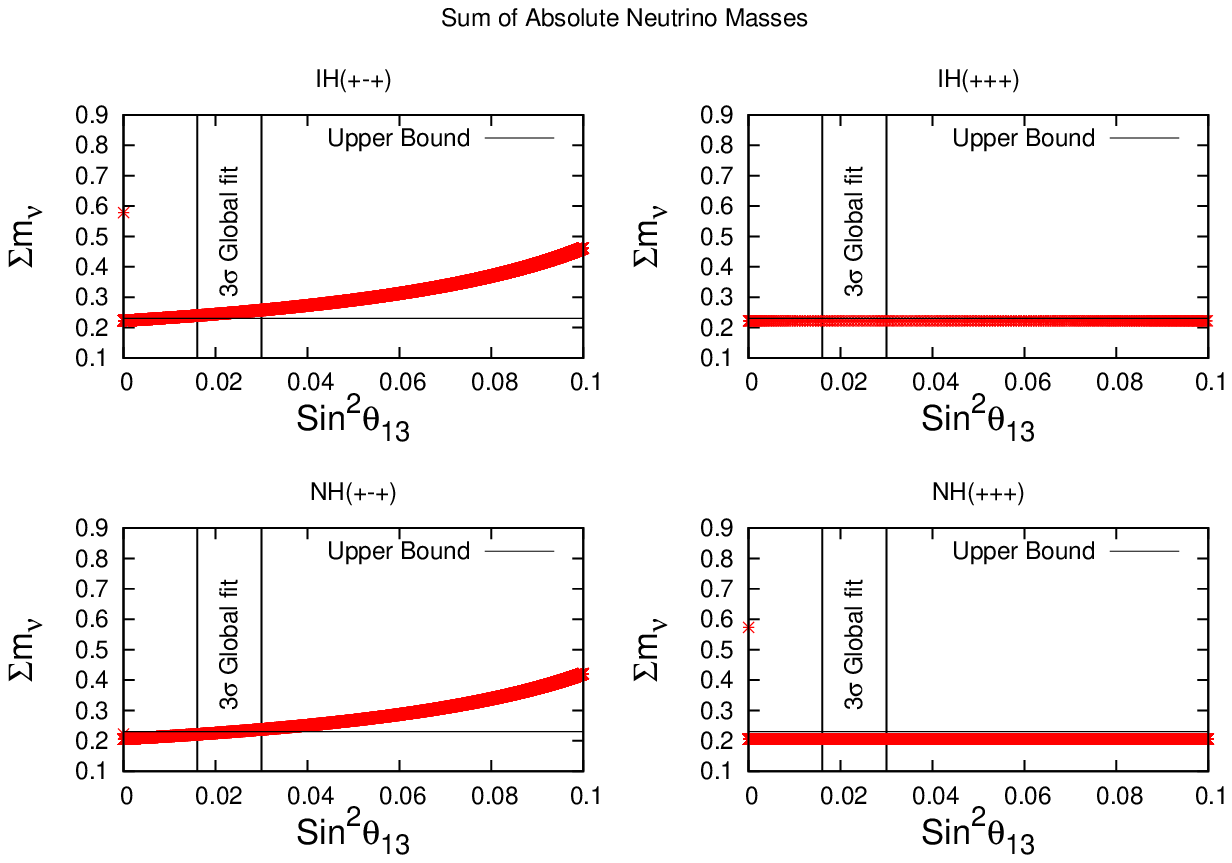}
\caption{Variation of the sum of absolute neutrino masses as a function of $\sin^2{\theta_{13}}$}
\label{fig5}
\end{figure}

After fitting the type I seesaw contribution to neutrino mass with experimental data, we introduce the type II seesaw contribution as a perturbation to the TBM neutrino mixing. Assuming the dimensionless couplings $f_{ij}$ to be of order one, the only free parameter we are left with in the type II seesaw term is $\langle \delta^0_L \rangle = v_L$, the vev acquired by the neutral component of $\Delta_L$. We compute the predictions for neutrino parameters by varying $v_L$ and show the results as a function of $\sin^2{\theta_{13}}$ in figure \ref{fig1}, \ref{fig2}, \ref{fig3} and \ref{fig4}. It should be noted that, although the type II seesaw term is introduced as a perturbation to TBM neutrino mass matrix in order to generate non-zero $\theta_{13}$, the other neutrino parameters also vary from their TBM values. As can be seen from the figures, only for the case of inverted hierarchy with $(+++)$ type Majorana phases, the other neutrino parameters lie within the $3\sigma$ range (marginally for $\Delta m^2_{21}, \sin^2{\theta_{23}}$) while keeping $\theta_{13}$ within $3\sigma$ range at the same time. Whereas for other models, the predictions for at least one of the neutrino parameters go outside the $3\sigma$ range for entire parameter space relevant to produce non-zero $3\sigma$ value of $\theta_{13}$. We also show the sum of absolute neutrino masses as a function of $\sin^2{\theta_{13}}$ in figure \ref{fig5} and find that they are consistent with the cosmological upper bound for all the models.

The variation of the neutrino parameters with the perturbation strength can be understood simply by calculating the diagonalizing matrix of the neutrino mass matrix considered in the study.
\begin{equation}
m_{LL}=\left(\begin{array}{ccc}
x& y-w&y+w\\
y-w& x+z-2w & y-z \\
y+w & y-z& x+z+2w
\end{array}\right)
\label{matrix3}
\end{equation}
which has eigenvalues $m_1 = x-y$, $m_2 = \frac{1}{2}(2x+y+2z-\sqrt{(3y-2z)^2+24w^2})$ and $m_3=\frac{1}{2}(2x+y+2z+\sqrt{(3y-2z)^2+24w^2})$. Assuming $m_1 < m_2 < m_3$ we calculate the neutrino parameters by first identifying the diagonalizing matrix. Assuming the perturbation strength $w$ to be very small such that $24w^2/(3y-2z)^2 \ll 1$ the dependence of neutrino mixing parameters on $w$ can be found as
\begin{equation}
\text{sin}^2{\theta_{13}} = \frac{\left(\frac{8z}{2z-3y}\right)^2w^2+\text{h.o.}}{16\left(2z^2-\frac{9(y+2z)}{2z-3y}w^2\right)+\text{h.o.}}
\end{equation}
\begin{equation}
\text{sin}^2{\theta_{12}} = \frac{\left(2y(2z-3y)-\frac{8z-24y}{2z-3y}w^2 \right)^2 +\text{h.o.}}{12((2z-3y)y-3yw+3y^2w)^2+\text{h.o.}}
\end{equation}
\begin{equation}
\text{sin}^2{\theta_{23}} = \frac{\left(4z-\frac{8z}{2z-3y}w \right)^2 +\text{h.o.}}{2\left(4z-\frac{8z(1-4z)}{2z-3y}w \right)^2+\text{h.o.}}
\end{equation}
where h.o. referes to higher order terms in $w$. It can be easily seen that for $w=0$, the mixing angles correspond to the values predicted by TBM mixing.

We can also constrain the value of $v_L$ in our model such that the predictions for neutrino parameters lie within the $3\sigma$ range. Considering the IH(+++) type model for which all the parameters are within or very close to the allowed range as can be seen from figure \ref{fig2}, we find the constraint on $v_L$ as
$$ v_L = 0.000047-0.000068 \;\; \text{eV} $$
which (from equation (\ref{vev})) would imply $M_{\Delta} \sim 10^{18}$ GeV which is very close to the reduced Planck scale $M_{Pl} \sim 2 \times 10^{18}$ GeV. If the origin of this extra Higgs triplet is associated with the GUT scale physics, then $M_{\Delta} \sim 10^{16}$ GeV, for which the dimensionless couplings $f_{ij}$ have to be fine-tuned by two orders of magnitudes in order to reproduce the correct $\theta_{13}$.

\section{Conclusion}
\label{conclude}
We have studied the possibility of explaining non-zero value of reactor mixing angle by introducing the type II seesaw term as a perturbation to the TBM type neutrino matrix derived from type I seesaw mechanism. Without considering the details of the origin of TBM mixing in type I seesaw models (which has been discussed many a times in the literature), we consider the minimal structure of the type II seesaw term to give the required perturbation to the TBM mixing. We justify this minimal choice of type II term by proposing two different flavor symmetry interpretations: abelian flavor symmetry and discrete flavor symmetry. We then numerically fit the type I seesaw term for TBM type $\mu-\tau$ symmetric neutrino mass matrix by taking oscillation data on mass squared differenes as well as cosmology data on the sum of absolute neutrino masses. We consider both normal and inverted hierarchical neutrino masses as well as two types of extremal Majorana phases denoted by $(+-+)$ and $(+++)$. After introducing the type II seesaw term as a perturbation, we compute the predictions for neutrino parameters as a function of the perturbation strength. We also compute the variation of the sum of absolute neutrino masses with this perturbation and check whether it stays below the cosmological upper bound. We find that only one scenario in our analysis (inverted hierarchy with $(+++)$ type Majorana phases) give rise to the correct $3\sigma$ values of all the neutrino parameters simultaneously. We then find the strength of the perturbation term for which the correct values of neutrino parameters are obtained. For order one dimensionless couplings, the Higgs triplet mass parameter $M_{\Delta}$ is found to be as high as $10^{18}$ GeV, in order to give rise to the correct perturbation strength. However, for slightly fine-tuned dimensionless couplings $f\sim 10^{-2}$, the origin of this triplet mass term can be related to the GUT scale $(\sim 10^{16}\;\text{GeV})$ physics. 

More precise experimental data should be able to shed more light on the viability of these models in giving rise to the correct phenomenology. It should be noted that, our analysis has not touched upon the important issue of Dirac neutrino phase which is taken to be zero here. Parametrizing the type II seesaw term by a complex parameter (instead of real as in the present analysis), it should be possible to predict the values of Dirac CP phase as well. Also, the Majorana phases (which are currently unconstrained from neutrino oscillation experiments) are assumed to take only extremal values in our analysis. A detailed analysis considering wider possibilities of these phases should allow more parameter space which are consistent with the experimental data. We leave such an exercise for future studies.


\end{document}